\documentclass[english]{article}
\usepackage[T1]{fontenc}
\usepackage[latin9]{inputenc}
\usepackage{geometry}
\usepackage{color}
\usepackage{array}
\usepackage{amstext}
\usepackage{graphicx}

\makeatletter

\providecommand{\tabularnewline}{\\}

\newcommand{\lyxaddress}[1]{
	\par {\raggedright #1
	\vspace{1.4em}
	\noindent\par}
}

\makeatother

\usepackage{babel}
\begin{document}
\title{Comment on ``Multi-output quantum teleportation of different quantum
information with an IBM quantum experience''}
\author{Satish Kumar \\ email: {satishphysics94@gmail.com} }
\maketitle

\lyxaddress{\begin{center}
Jaypee Institute of Information Technology, A 10, Sector 62, Noida,
UP-201309, India
\par\end{center}}
\begin{abstract}
Recently, Yu et al., (Commun. Theor. Phys. \textbf{73} (2021) 085103)
has proposed a scheme for \textquotedbl multi-output quantum teleportation''
and has implemented the scheme using an IBM quantum computer. In their
so called multicast-based quantum teleportation scheme, a sender (Alice)
teleported two different quantum states (one of which is a $m$-qubit
GHZ class state and the other is a $(m+1)$-qubit GHZ class state)
to the two receivers. To perform the task, a five-qubit cluster state
was used as a quantum channel, and the scheme was realized using IBM
quantum computer for $m=1$. In this comment, it is shown that the
quantum resources used by Yu et al., was extensively high. One can
perform the same task of two-party quantum teleportation using two
Bell states only. The modified scheme for multi-output teleportation
using optimal resources has also been implemented using IBM quantum
computer for $m=1$ and the \textcolor{black}{obtained result is
compared with the result of Yu et al. }
\end{abstract}

\section{Introduction}

Quantum teleportation scheme was first introduced by Bennett et al.
in 1993. In their pioneering work, they proposed a scheme for teleporting
an unknown qubit using a maximally entangled Bell state \cite{bennett1993}.
Since then many teleportation schemes have been proposed and many
variants of teleportation (e.g., remote state preparation (RSP), quantum
secret sharing (QSS), quantum information splitting (QIS), bidirectional
teleportation (see \cite{pathak2013elements,sharma2015controlled}
and references therein)) have also been introduced. Recently Yu et
al. \cite{yu2017} have proposed a scheme for the teleportation of
two different quantum states to two different receivers. Specifically,
in their scheme, Alice wants to teleport a state

\begin{eqnarray}
|\chi_{a}\rangle & = & \alpha_{1}|0\rangle^{\otimes m}+\beta_{1}|1\rangle^{\otimes m}\label{eq:1}
\end{eqnarray}
 and an another state

\begin{eqnarray}
|\chi_{b}\rangle & = & \alpha_{2}|0\rangle^{\otimes(m+1)}+\beta_{2}|1\rangle^{\otimes(m+1)}\label{eq:2}
\end{eqnarray}
 to Bob$_{1}$ and Bob$_{2}$, respectively using a five qubit-cluster
state

\begin{eqnarray}
|\psi\rangle_{12345} & = & \frac{1}{2}(|00000\rangle+|01011\rangle+|10100\rangle+|11111\rangle)_{12345}.\label{eq:3}
\end{eqnarray}

In their work, Yu et al. have referred to state $|\chi_{a}\rangle$
as $m$-qubit state of GHZ class and analogously $|\chi_{b}\rangle$
as $(m+1)$-qubit state of GHZ class. We have some reservation about
using this nomenclature and would prefer to name these states as generalized
Bell-type state as was logically done in several earlier works \cite{pathak2011,panigrahi2006}.
In fact, in Ref. \cite{pathak2011}, it was explicitly shown that
any quantum state of the from $\alpha|x\rangle+\beta|\bar{x}\rangle:|\alpha|^{2}+|\beta|^{2}=1$
where $x$ varies from $0$ to $2^{n}-1$ and $\bar{x}=1^{\otimes n}\oplus x$
in modulo 2 arithmetic, can be teleported using a Bell state. Clearly,
the states considered by Yu et al. (i. e., $|\chi_{a}\rangle$ and
$|\chi_{b}\rangle$) are of the form $\alpha|x\rangle+\beta|\bar{x}\rangle$
and it's obvious that $|\chi_{a}\rangle$ and $|\chi_{b}\rangle$
can be independently teleported to two receivers using two Bell states.
Thus, the use of a five qubit-cluster state or any such complicated
quantum channel is not required to perform the multi-output teleportation
task considered by Yu et al. 

Extending the above observation, it will be apt to note that a generalized
scheme for teleportation has been reported in \cite{sisodia2017},
where it is mentioned that teleportation of a quantum state having
$m$-unknown coefficients require $\lceil\log_{2}m\rceil$ Bell states.
The scheme proposed by Yu et al., is essentially meant for teleportation
of a product state $|\psi_{ab}\rangle=|\chi_{a}\rangle\otimes|\chi_{b}\rangle$
having four unknown coefficient $\alpha_{1}$, $\beta_{1}$, $\alpha_{2}$,
and $\beta_{2}$ and hence require only $\lceil\log_{2}4\rceil=2$
Bell states to perform the teleportation task. In fact, the scheme
of \cite{sisodia2017} allows one to teleport more general quantum
states using two Bell state\textcolor{black}{s}. Interestingly, despite
the existence of these general results, several authors have recently
reported different type of teleportation schemes using excessively
higher amount of quantum resource. For example, in Ref. \cite{bikash2020}
a four qubit cluster state is used as a quantum resource for teleporting
two qubit\textcolor{black}{{} states}. The two qubit \textcolor{black}{states}
used for teleportation is

\begin{eqnarray}
|\lambda\rangle_{ab} & = & \alpha|00\rangle+\beta|01\rangle+\gamma|10\rangle+\delta|11\rangle.\label{eq:4}
\end{eqnarray}
Now, as per the scheme reported in \cite{sisodia2017} and the references
therein, since there are four unknown coefficients in the state to
be teleported, it will be sufficient to use$\lceil\log_{2}4\rceil=2$
Bell states. The discussion so far is sufficient to establish that
the resources used in Yu et al., paper are not optimal and we could
have concluded this comment here, but the fact that they have realized
their scheme for $m=1$ using IBM quantum experience have motivated
us to explicitly implement our scheme scheme for $m=1$ with the help
of a quantum computer whose cloud-based access is provided by IBM.

The paper is organized as follows. Our scheme for multi-output teleportation
using two Bell states is described in Sec. \ref{sec:circui}. Subsequently,
the implementation of the scheme using an IBM quantum computer and
the relevant results are reported in Sec. \ref{sec:Experimental-realization-usingIBM}.
Finally, the paper is concluded in Sec. \ref{sec:Conclusion}. 

\section{Multi-output quantum teleportation using two Bell states\label{sec:circui}}

\begin{figure}
\begin{centering}
\includegraphics[scale=0.5]{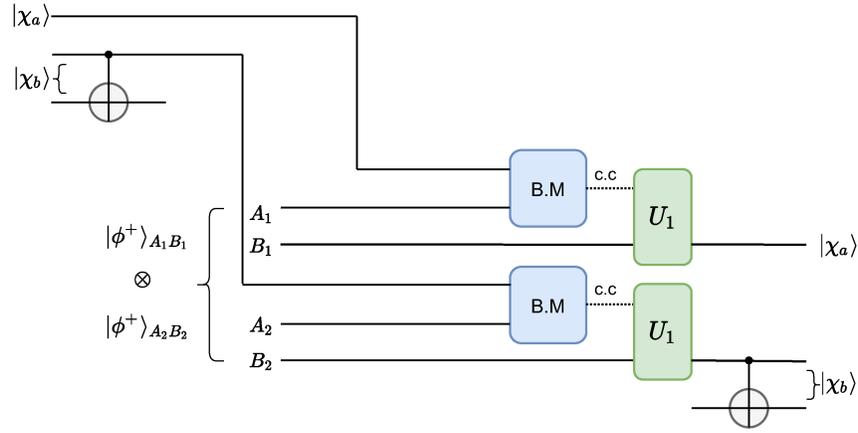}
\par\end{centering}
\caption{(Color online) An optimal quantum circuit illustrating a multi-output
quantum teleportation scheme\label{fig:Multi-output-quantum-teleportati}}

\end{figure}

In 2017, Yu et al., coined the term multi-output quantum teleportation
{\cite{yu2017}} in an effort to propose a scheme that
allows Alice to teleport two different single qubit states $|\chi_{1}\rangle$and
$|\chi_{2}\rangle$ to two different receivers using a four qubit
cluster state $|\psi\rangle_{A_{1}A_{2}B_{1}B_{2}}$. In the original
scheme, Alice used to keep the first two qubits (indexed by subscripts
$A_{1}$and $A_{2}$) of the cluster state with herself and sends
the other two qubits to the two receivers, say Bob$_{1}$ and Bob$_{2}$
(qubits sent to ${\rm Bob}_{i}$ is indexed by $B_{i}$). Now, Alice
does a measurement in the cluster basis on first four qubits $|\psi_{i}\rangle_{12A_{1}A_{2}}$,
two of which are information qubits and the other two are the qubits
which Alice kept with her. The measurement result is publicly announced
and the two receivers applies the corresponding unitary operators
to obtain the corresponding desired states $|\chi_{1}\rangle$ and
$|\chi_{2}\rangle$, respectively. Almost in the similar line, in
2021, Yu et al., proposed another scheme for multi-output quantum
teleportation, but this time the states to be teleported were $m$-qubit
and $(m+1)$-qubit states (cf. Eqs. (\ref{eq:1}) and (\ref{eq:2})
and the related discussions in the previous section) and the quantum
channel used was a five-qubit cluster state (see Eq. \ref{eq:3}).
We have already mentioned that the same multi-output teleportation
task can be done using two bell states. As the experimental part of
Yu et al. is restricted to $m=1$ case, for comparison, in Fig. \ref{fig:Multi-output-quantum-teleportati},
we explicitly show the schematic of the quantum circuit that will
be required for performing the task using two Bell states. Let $|\chi_{a}\rangle$
and $|\chi_{b}\rangle$ be the two states to be teleported (Eqs. (\ref{eq:1})
and (\ref{eq:2}) for $m=1)$. The state $|\chi_{b}\rangle$ can be
reduced to a simpler state $|\chi_{b}^{\prime}\rangle$ after applying
a unitary operation CNOT with control on first qubit and target on
second qubit. Now the problem reduces to the teleportation of the
product state of $|\chi_{a}\rangle$ and |$\chi_{b}^{\prime}\rangle=\alpha_{2}|0\rangle+\beta_{2}|1\rangle$
as

\begin{eqnarray}
CNOT|\chi_{b}\rangle & \longrightarrow & |\chi'_{b}\rangle|0\rangle=(\alpha_{2}|0\rangle+\beta_{2}|1\rangle)\otimes|0\rangle.\label{eq:5}
\end{eqnarray}

\section{Experimental realization using an IBM quantum computer\label{sec:Experimental-realization-usingIBM}}

\begin{figure}
\begin{centering}
\includegraphics[scale=0.5]{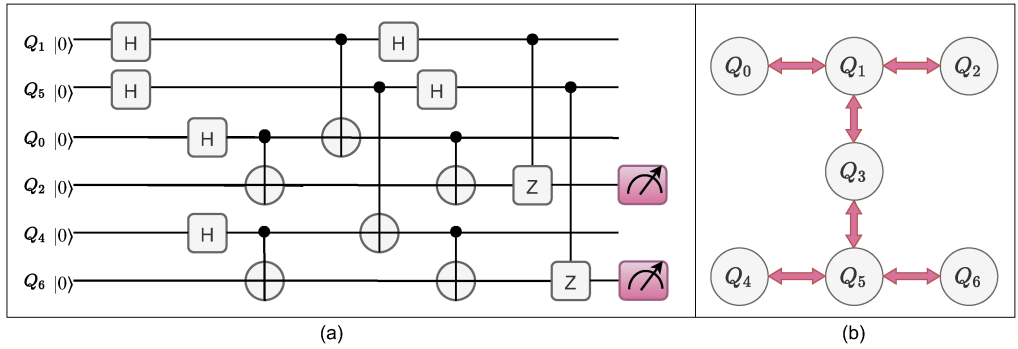}
\par\end{centering}
\centering{}\caption{(Color online) (a) Quantum circuit for the teleportation of the states
$\frac{1}{\sqrt{2}}(|0\rangle+|1\rangle)$ and $\frac{1}{\sqrt{2}}(|0\rangle+|1\rangle)$
to two different receivers Bob$_{1}$ and Bob$_{2}$ simultaneously
using 2 Bell states $|\phi^{+}\rangle^{\otimes2}$ (b) Topology of
ibmq\_casablanca. \label{fig:MQTCkt}}
\end{figure}

We have designed a simple (but experimentally realizable using IBM
quantum experience) circuit shown in Fig. \ref{fig:MQTCkt} (a) which
is equivalent to the schematic of the circuit shown in Fig. \ref{fig:Multi-output-quantum-teleportati}
except the presence of the first and last CNOT gates. Local operations
performed by these two CNOT gates do not affect the main teleportation
part. This circuit is run in IBM quantum composer to yield the results
reported in the following subsection. There is another reason for
implementing the circuit without the CNOT gates, as that allowed us
to use ibmq\_casablanca which is a seven qubit quantum computer that
has enough resources to implement the circuit shown in Fig. \ref{fig:MQTCkt}
(a), but not enough qubits to implement the technically equivalent
circuit shown in Fig. \ref{fig:Multi-output-quantum-teleportati}.
The ibmq\_casablanca is one of the IBM Quantum Falcon processors \cite{ibm2021}.
The circuit given in Fig. \ref{fig:MQTCkt} (a) can be briefly described
as a process in which Alice wants to teleport $\text{\ensuremath{\frac{1}{\sqrt{2}}}(|0\ensuremath{\rangle}+|1\ensuremath{\rangle})\ensuremath{\otimes\ensuremath{\frac{1}{\sqrt{2}}}(|0\ensuremath{\rangle}+|1\rangle)=|+\rangle\otimes|+\rangle}}$
to the receivers Bob$_{1}$ and Bob$_{2}$, respectively as the first
CNOT in Fig. \ref{fig:Multi-output-quantum-teleportati} can transform
the Bell state $|\phi^{+}\rangle=\frac{1}{\sqrt{2}}(|00\rangle+|11\rangle)$
to separable state $\frac{1}{\sqrt{2}}(|0\rangle+|1\rangle)\otimes|0\rangle$
and the last CNOT can recreate it at the receiver's end with the help
of an ancilla qubit and output of the teleportation process. In Fig.
\ref{fig:MQTCkt} (a), first two qubits are the information qubits
and the last four qubits are the quantum channels used for the teleportation,
which is comprised of two Bell states as desired and argued above
as sufficient resource. Now Alice does a Bell measurement on first
($Q_{1}$) and third qubits ($Q_{0}$) and another Bell measurement
on second ($Q_{5}$) and fifth qubits ($Q_{4}$) and then sends the
measurement results to Bob$_{1}$ and Bob$_{2}$. Here it may be noted
that qubit numbers are indexed in accordance with the convention adopted
by IBM Quantum experience in describing the 7 qubit quantum computer
whose topology is shown in Fig. \ref{fig:MQTCkt} (b). Further, the
qubits are chosen such that the circuit after transpilation has a
minimal circuit cost \cite{dueck2018optimization}. According to the
measurement results announced by Alice, Bob$_{1}$ and Bob$_{2}$
apply corresponding unitaries to obtain the teleported states. Clearly
this is just two independent implementation of the standard teleportation
circuit, and the same is enough to achieve what is done using costly
quantum resources in the earlier works.

\subsection{Results}

\begin{figure}
\begin{centering}
\includegraphics[scale=0.6]{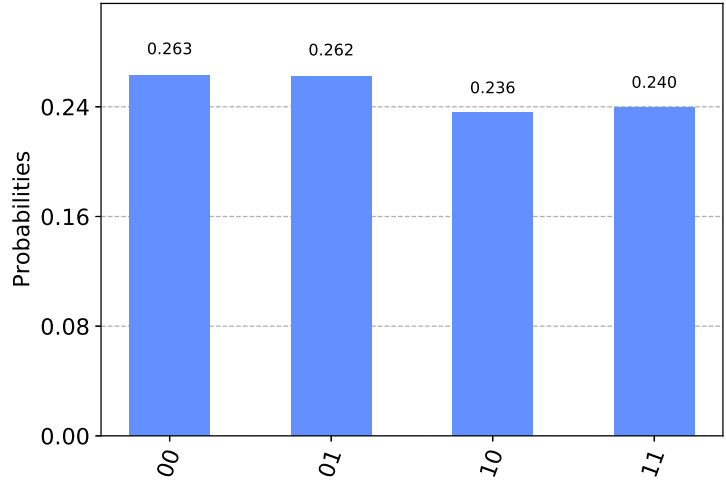}
\par\end{centering}
\centering{}\caption{(Color online) Experimental result for the quantum circuit shown in
Fig \ref{fig:MQTCkt}. \label{fig:results}}
\end{figure}

\begin{figure}
\begin{centering}
\includegraphics[scale=0.6]{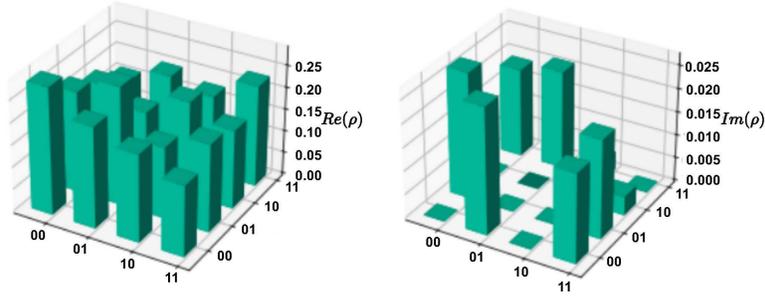}
\par\end{centering}
\caption{(Color Online) Experimental quantum state tomography result for the
circuit shown in Fig \ref{fig:MQTCkt}.\label{fig:ExDM}}
\end{figure}

The circuit described above is run using ibmq\_casablanca which is
a 7 qubit superconductivity based quantum computer that uses transmon
qubits. The obtained result is illustrated in Fig \ref{fig:results}.
As we teleported $|+\rangle|+\rangle$ it was expected that in output
states $|00\rangle,|01\rangle,|10\rangle$ and $|11\rangle$ would
appear with equal probability, but from Fig. \ref{fig:results} we
can see that the states are produced with slightly different probabilities,
the same is also depicted in the corresponding density matrix shown
in Fig. \ref{fig:ExDM}. This is because of the inherent implementation
errors as summarized in Table \ref{tab:Calibration-data-of}. Fidelity
between the state produced and the expected state is computed using
the formula $F(\sigma,\rho)=Tr\left[\sqrt{\sqrt{\sigma}\rho\sqrt{\sigma}}\right]^{2}$,
where $\sigma$ is theoretical (expected) density matrix of the final
state and $\rho$ is the density matrix of the experimentally obtained
final state. The fidelity is obtained as 84.64 \% for the case illustrated
here for a particular set of experiment comprised of 8192 runs of
the experiment. To check the consistency of the result the same exercise
is repeated 10 times and the fidelities are obtained as (in \%) 77.51,
84.64, 79.31, 78.98, 76.17, 81.33, 83.64, 80.21, 74.65, 79.92. The
standard deviation is 3.096. This is a reasonably accurate result
and the fidelity is quite high compared to the classical limit of
2/3. This simply establishes that resources used in the earlier works
were not optimal. Fidelity can not be compared with the earlier work,
as Yu et al., have not reported that. However, it's obvious that simpler
entangled states used here will be \textcolor{black}{affected} less
by the noise. 

\begin{table}
\begin{centering}
\begin{tabular}{|c|c|c|>{\centering}p{2cm}|>{\centering}p{2cm}|>{\centering}p{2cm}|>{\centering}p{5cm}|}
\hline 
Qubit & T1 ($\mu s$) & T2 ($\mu s$) & \centering{}Frequency (GHz) & \centering{}Readout assignment error & \centering{}Single-qubit Pauli-X-error & \centering{}CNOT error\tabularnewline
\hline 
$Q_{0}$ & 97.07 & 41.56 & \centering{}4.822 & \centering{}$3.52\times10^{-2}$ & \centering{}$2.73\times10^{-4}$ & \centering{}cx0\_1: $1.105\times10^{-2}$\tabularnewline
\hline 
$Q_{1}$ & 179.27 & 106.63 & \centering{}4.76 & \centering{}$1.56\times10^{-2}$ & \centering{}$1.56\times10^{-4}$ & \centering{}cx1\_3: $6.796\times10^{-3}$, cx1\_2: $1.013\times10^{-2}$,
cx1\_0: $1.105\times10^{-2}$\tabularnewline
\hline 
$Q_{2}$ & 164.86 & 96.43 & \centering{}4.906 & \centering{}$8.50\times10^{-3}$ & \centering{}$3.54\times10^{-4}$ & \centering{}cx2\_1: $1.013\times10^{-2}$\tabularnewline
\hline 
$Q_{3}$ & 123.23 & 151.27 & \centering{}4.879 & \centering{}$1.70\times10^{-2}$ & \centering{}$3.40\times10^{-4}$ & \centering{}cx3\_1: $6.796\times10^{-3}$, cx3\_5: $1.139\times10^{-2}$\tabularnewline
\hline 
$Q_{4}$ & 128.4 & 54.14 & 4.871 & \centering{}$3.06\times10^{-2}$ & \centering{}$2.88\times10^{-4}$ & cx4\_5: $1.148\times10^{-2}$\tabularnewline
\hline 
$Q_{5}$ & 133.5 & 91.77 & 4.964 & \centering{}$9.60\times10^{-3}$ & \centering{}$3.17\times10^{-4}$ & cx5\_3: $1.139\times10^{-2}$, cx5\_4:$1.148\times10^{-2}$, cx5\_6:
$1.156\times10^{-2}$,\tabularnewline
\hline 
$Q_{6}$ & 112.08 & 166.07 & 5.177 & \centering{}$2.18\times10^{-2}$ & \centering{}$4.70\times10^{-4}$ & cx6\_5:$1.156\times10^{-2}$\tabularnewline
\hline 
\end{tabular}
\par\end{centering}
\caption{Calibration data of ibmq\_casablanca on Dec 01, 2021.cxi\_j represents
CNOT gate with control qubit i and target qubit j.\label{tab:Calibration-data-of}}
\end{table}

\section{Conclusion\label{sec:Conclusion}}

It's shown that quantum resources used in Yu et al., \cite{yu2021}
for multiparty teleportation was not optimal and the same drawback
exists in \cite{bikash2020} and other similar works. Relevant existing
results are noted and it's explicitly shown that ibmq\_casablanca
can be used to implement the task described by Yu et al., using only
two Bell states. Here the purpose was only to show that cluster state
and similar resources are not required for performing this type of
tasks, and consequently we have restricted ourselves to simplest possible
implementation of the multi-output quantum teleportation. It's obvious
to extend this approach for the multioutput teleportation of more
complex quantum states.

\section*{Acknowledgment:}

Author acknowledges the support from the QUEST scheme of Interdisciplinary
Cyber Physical Systems (ICPS) program of the Department of Science
and Technology (DST), India (Grant No.: DST/ICPS/QuST/Theme-1/2019/14
(Q80)). He also thanks Anirban Pathak for his feedback and advises
in relation to the present work.

\bibliographystyle{ieeetr}
\bibliography{multioutput_ref}

\end{document}